\documentclass{iau}
\usepackage{graphicx, bm, amssymb, xcolor}
\usepackage{aas_macros}
\usepackage{hyperref}

\usepackage{verbatim}
\usepackage[all]{hypcap}
\usepackage{xspace}
\usepackage{multirow}

\usepackage{epsfig}
\usepackage{natbib}

\title[Gas and star formation in the CMZ]
{Towards a multi-scale understanding\\of the gas-star formation cycle\\in the Central Molecular Zone}

\author[J.~M.~Diederik Kruijssen]
{J.~M.~Diederik Kruijssen}

\affiliation{Astronomisches Rechen-Institut, Zentrum f\"{u}r Astronomie der Universit\"{a}t Heidelberg, M\"{o}nchhofstra\ss e 12-14, 69120 Heidelberg, Germany \\ email: {\tt kruijssen@uni-heidelberg.de}}

\pubyear{2016}
\volume{322}
\pagerange{}

\jname{Astrophysics of the Galactic Centre}
\editors{R.~Crocker, S.~Longmore \& G.~Bicknell}

\begin{document}

\maketitle

\begin{abstract}
The Central Molecular Zone (CMZ, the central 500 pc of the Milky Way) contains the largest reservoir of high-density molecular gas in the Galaxy, but forms stars at a rate 10--100 times below commonly-used star formation relations. We discuss recent efforts in understanding how the nearest galactic nucleus forms its stars. The latest models of the gas inflow, star formation, and feedback duty cycle reproduce the main observable features of the CMZ, showing that star formation is episodic and that the CMZ currently resides at a star formation minimum. Using orbital modelling, we derive the three-dimensional geometry of the CMZ and show how the orbital dynamics and the star formation potential of the gas are closely coupled. We discuss how this coupling reveals the physics of star formation and feedback under the conditions seen in high-redshift galaxies, and promotes the formation of the densest stellar clusters in the Galaxy.
\end{abstract}

\firstsection

\section{Introduction} \label{sec:intro}
The gas-star formation cycle near the Galactic Centre is inherently a multi-scale process. There exists a close interplay between large-scale gas flows, galactic dynamics, star formation, feedback, and the feeding of the central supermassive black hole, Sgr~A$^*$. The dominant mass reservoir containing the fuel for star formation near the Galactic Centre is the Central Molecular Zone (CMZ; i.e.~the central 500 pc of the Milky Way). This region contains the largest concentration of high-density molecular gas in the Galaxy ($M_{\rm gas}\sim5\times10^7~{\rm M}_\odot$, \citealt{ferriere07}) and obtaining an understanding of the gas-star formation cycle in the CMZ is not just valuable from a star formation perspective, but also has key implications for wider areas in astrophysics and Galactic Centre research.

There are several examples of the wider implications of CMZ studies. For instance, the physics driving the accretion and activity of Sgr~A$^*$ govern a large range of spatial scales with several independent bottlenecks. How is the gas deposited into the CMZ from the Galactic disc? Which transport mechanisms drive it further inwards once it has passed the inner Lindblad resonance (ILR)? How does it reach the sphere of influence of the nuclear cluster? How does it eventually reach the accretion disc of Sgr~A$^*$? The large-scale flow plays a critical role in providing the material for the eventual activity of Sgr~A$^*$.

A related, major example of the importance of the gas in the CMZ is its three-dimensional geometry. How does the distribution of absorbers along the line of sight affect the ongoing search for dark matter annihilation signals from the Galactic Centre \citep[e.g.][]{daylan16}? A three-dimensional model for the gas in the CMZ would also provide a detailed record of the recent ($<10^3$ years) accretion history of Sgr~A$^*$ from X-ray light echoes \citep[e.g.][and these proceedings]{clavel13,clavel14}. While the accretion history over much longer time-scales may be reconstructed with high-energy observations of the outflowing relics in the Galactic halo, there exists an important degeneracy with feedback from (possibly bursty) star formation \citep[e.g.][]{su10}. If we can achieve an understanding of where and how stars from in the CMZ, this may allow this degeneracy to be lifted. At the same time, such an understanding will yield insights in the growth and structural evolution of the central regions of the bulge, where the baryons dominate the gravitational potential by orders of magnitude. Finally, the densest stellar clusters in the Milky Way are situated within 100~pc of the Galactic Centre (\citealt{walker16}; including the nuclear cluster, e.g.~\citealt{genzel10b}). The close proximity of the CMZ provides a unique opportunity for studying how such extreme stellar populations are formed, as well as for constraining the formation rates and mechanisms of compact objects within these clusters.

Below, we describe our first steps in synthesising the multi-scale structure of the gas-star formation cycle in the CMZ, from the inflow of fresh gas along the bar (of the order $1~{\rm M}_\odot~{\rm yr}^{-1}$) to the accretion onto the supermassive black hole in Sgr~A* (of the order $10^{-9}~{\rm M}_\odot~{\rm yr}^{-1}$). The end goal is to obtain a complete understanding of the structure and evolution of the CMZ, which will help answering many of the questions raised above.

\section{Macro-evolution of the gas-star formation cycle in the CMZ} \label{sec:macro}
\subsection{A low star formation rate in the CMZ}
Despite the large reservoir of dense gas ($M_{\rm gas}\sim5\times10^7~{\rm M}_\odot$) in the CMZ, it forms stars at a rate of ${\rm SFR}=0.05$--$0.15~{\rm M}_\odot~{\rm yr}^{-1}$ \citep[e.g.][]{yusefzadeh09,immer12,barnes16}, which is 10-100 times lower than expected based on commonly-used star formation relations \citep{longmore13} or the predictions of turbulent star formation theory \citep[e.g.][]{krumholz05,padoan11}. The underproduction of stars is particularly surprising in view of the gas properties -- the CMZ is characterised by gas pressures, velocity dispersions, and densities that are orders of magnitude higher than those in the solar neighbourhood, but similar to those in high-redshift galaxies \citep{kruijssen13b}. These empirical, macroscopic properties of the gas and star formation content suggest that the gas-star formation balance in the CMZ is unlikely to remain static over multiple dynamical times.

There are three possible explanations for the low SFR in the CMZ. Firstly, it could be an observational error due to systematics in the conversions from gas and star formation tracer emission to physical gas masses and SFRs. Secondly, it may be that there is something about the gas in the CMZ that makes it consistently inefficient over long time-scales (e.g.~a large number of dynamical times). Thirdly, it could be that the SFR in the CMZ undergoes an (episodic) time-variation, in which case the CMZ may currently reside near a star formation minimum. The first of these explanations has recently been ruled out by \citet{barnes16}, who find that a broad range of independent gas and star formation tracers all yield gas masses and SFRs that agree to within a factor of $\sim2$.

\subsection{A self-consistent cycle of episodic star formation}
In \citet{kruijssen14b}, we carried out a systematic study of all mechanisms we considered could conceivably suppress the SFR in the CMZ over long time-scales, such as extreme stability of the gas disc, cloud disruption by tidal perturbations, galactic shear, extreme turbulence, magnetic fields, radiation pressure feedback, or cosmic rays (either from star formation or from Sgr A$^*$). We concluded that none of these mechanisms could inhibit star formation for more than 5--10 Myr and therefore proposed that the reason for the {\it presently}-low SFR in the CMZ is that it undergoes an episodic cycle of gas inflow, star formation, and feedback. This suggestion is corroborated by observed signs of past starburst activity in the CMZ. The region hosts a large population of 24$\mu$m sources at negative longitudes \citep[e.g.][]{bally10} that may have formed during a previous star formation episode, as well as the two young, massive Arches and Quintuplet clusters \citep[e.g.][]{longmore14}, and the CMZ also resides at the root of the Fermi bubbles, which are possibly star formation-driven \citep[e.g.][]{su10,carretti13}.

Further evidence for nuclear episodicity is provided by star formation studies in extragalactic centres. \citet{leroy13} find that, while the {\it median} gas depletion time ($t_{\rm depl}=M_{\rm gas}/{\rm SFR}$) decreases towards galactic centres (i.e.~star formation gets more efficient on average), the scatter of the gas depletion time also greatly increases. In principle, this increase of the scatter could arise if there exist larger systematic differences in gas or star formation properties between galactic centres than between galactic discs, but an equally plausible explanation is that over time, galactic centres undergo larger excursions in their gas depletion time than galactic discs.

\begin{figure}
\vspace{-1.2cm}
\includegraphics[width=9.55cm, angle=-90]{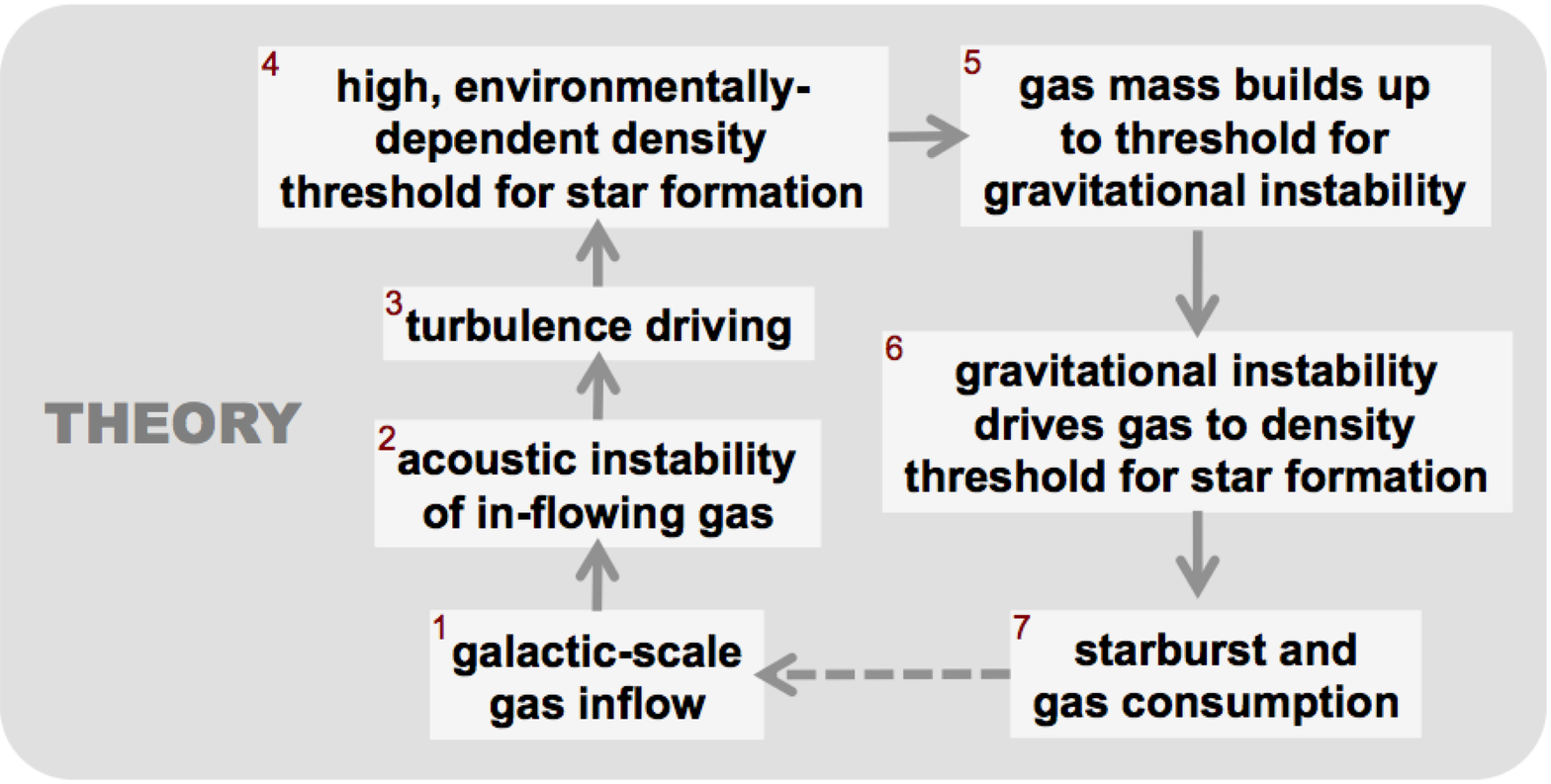}
\vspace{-1.2cm}
\caption{
\label{fig:schem}
Schematic representation of the star formation cycle in the CMZ proposed by \citet{kruijssen14b}, which we further developed and quantified in \citet[boxes 1--5]{krumholz15} and \citet[the full cycle]{krumholz16}. The arrow connecting box 7 to box 1 is dashed, because it only indicates the progression of time, whereas the other arrows follow the evolution of a single mass element. Adapted from Figure~6 in \citet{kruijssen14b}.
}
\end{figure}
Specifically, \citet{kruijssen14b} proposed a scenario for episodic star formation in the CMZ that is schematically illustrated in Figure~\ref{fig:schem}. The torques exerted by the bar drive a large-scale gas inflow to within the ILR (box~1), which in the Milky Way resides at $\sim1$~kpc \citep{krumholz15} and where density waves form from pressure-driven acoustic instabilities (box~2, \citealt{montenegro99}). Contrary to density waves outside the ILR, the acoustic instability is maintained by the repulsive force from the pressure gradients of the waves, whereas the attractive force from self-gravity erases the instability. The density waves are thus stable to gravitational collapse and can consistently drive turbulent energy into the gas (box~3), which in turn increases the critical (over)density threshold for star formation \citep[box~4, e.g.][]{krumholz05,padoan11}. Over time, this highly stable gas reservoir grows, until it finally reaches a high enough density to become gravitationally unstable (box~5). As the virial ratio of the gas decreases towards self-gravity, gas overcomes the density threshold for star formation (box~6), which at the high densities ($\sim10^4$~cm$^{-3}$) and short free-fall times ($<0.5$~Myr) of the CMZ leads to rapid star formation (box~7). The feedback from the young stellar population eventually drives down the star formation rate and the cycle starts anew.

\subsection{A dynamical model for star formation in the CMZ} \label{sec:model}
In \citet{krumholz15} and \citet{krumholz16}, we develop a dynamical model for episodic star formation and feedback in CMZs that quantifies the scenario of Figure~\ref{fig:schem} further. The former of these papers focuses on boxes 1--5 and in particular on modelling the acoustic instability. We describe a one-dimensional hydrodynamical disc model, with an outer boundary condition set by the gas inflow rate. Further radial inflow is caused by shear-driven angular momentum transport by acoustic instabilities. The transport drives turbulence and keeps the gas temporarily highly gravitationally stable. Because turbulence dissipates on a vertical crossing time ($<1$~Myr), collapse takes place almost immediately once the turbulence driving mechanism shuts off. In the context of this model, that driving mechanism is shear, and a key prediction of the model is therefore that the dynamical state of molecular clouds in CMZs is closely linked to the shape of the rotation curve.

One of the main quantitative predictions of the \citet{krumholz15} model for the Galactic CMZ is that at large radii ($R=150$--$500$~pc), the gas is inflowing at $v_R\sim-20$~km~s$^{-1}$, with a low surface density ($\Sigma\sim30$~M$_\odot$~pc$^{-2}$) and a large linewidth ($\sigma_{1{\rm D}}\sim30$~km~s$^{-1}$), rendering it highly gravitationally stable ($Q_{\rm gas}\sim100$) with a low star formation efficiency. However, as the gas is flowing in, the rotation curve eventually turns over from flat to being close to solid body. This happens at a radius of $R\sim100$~pc, where the shear drops steeply, the inflow stalls, the gas density increases, the turbulent energy dissipates, and the clouds become gravitationally unstable and prone to star formation. This dichotomy in gas properties between $R\sim100$~pc and larger radii is consistent with observed differences in the total gas mass, the gas density and scale height, and most importantly the virial parameters of the clouds \citep{kruijssen14b,walker15}.

The model is expanded by including a simple description of star formation and feedback in \citet{krumholz16}, where we thus add boxes~6 and~7 in Figure~\ref{fig:schem} and close the cycle. This addition enables the modelling of multiple cycles as well as feedback-driven galactic winds. As discussed in the paper, we indeed find that star formation in the CMZ is episodic, with a characteristic period of 10--20~Myr. However, contrary to the original expectation of box~7 in Figure~\ref{fig:schem}, the starburst is not truncated by gas consumption or removal, but by the large increase of the turbulent energy in the gas due to stellar feedback. This puts the clouds back in a highly supervirial state, allowing the cycle to restart from the beginning. Interestingly, the \citet{krumholz16} model does not only reproduce the current Galactic CMZ, but its predicted time-evolution also provides a good match to the distribution of gas depletion times observed in extragalactic CMZs.

As described, the qualitative scenario of Figure~\ref{fig:schem} has now matured to a first quantitative model that describes the macroscopic gas-star formation cycle in CMZs. Out of the three possible explanations for the low observed SFR in the CMZ, the evidence is now clearly pointing to the third option: the SFR in the CMZ undergoes an (episodic) time-variation and the CMZ currently resides near a star formation minimum.

\section{Origin, dynamical evolution and 3D structure of the 100-pc stream}
\subsection{Orbital structure}
As discussed in \S\ref{sec:macro}, the transition of the Galactic rotation curve from flat at large radii to near-solid body at $R\sim100$~pc is predicted to result in a pile-up of the inflowing gas at that radius due to the shear minimum in the Galactic rotation curve. By accumulating at this radius, the gas can finally become gravitationally unstable and undergo widespread star formation. Indeed, \citet{molinari11} have reported the existence of such a ring-like gas stream with a radius of $R\sim100$~pc. The observation of \citet{molinari11} was followed up by \citet{kruijssen15}, who calculated orbital models in the gravitational potential of the central Milky Way \citep{launhardt02} to fit the $\{l,b,v_{\rm los}\}$ structure of the gas in the CMZ. As discussed in \S\ref{sec:intro}, having a three-dimensional model of the gas in the CMZ is a key goal in Galactic Centre physics.
\begin{figure}
\center
\vspace{-1.2cm}
\includegraphics[width=6.44cm]{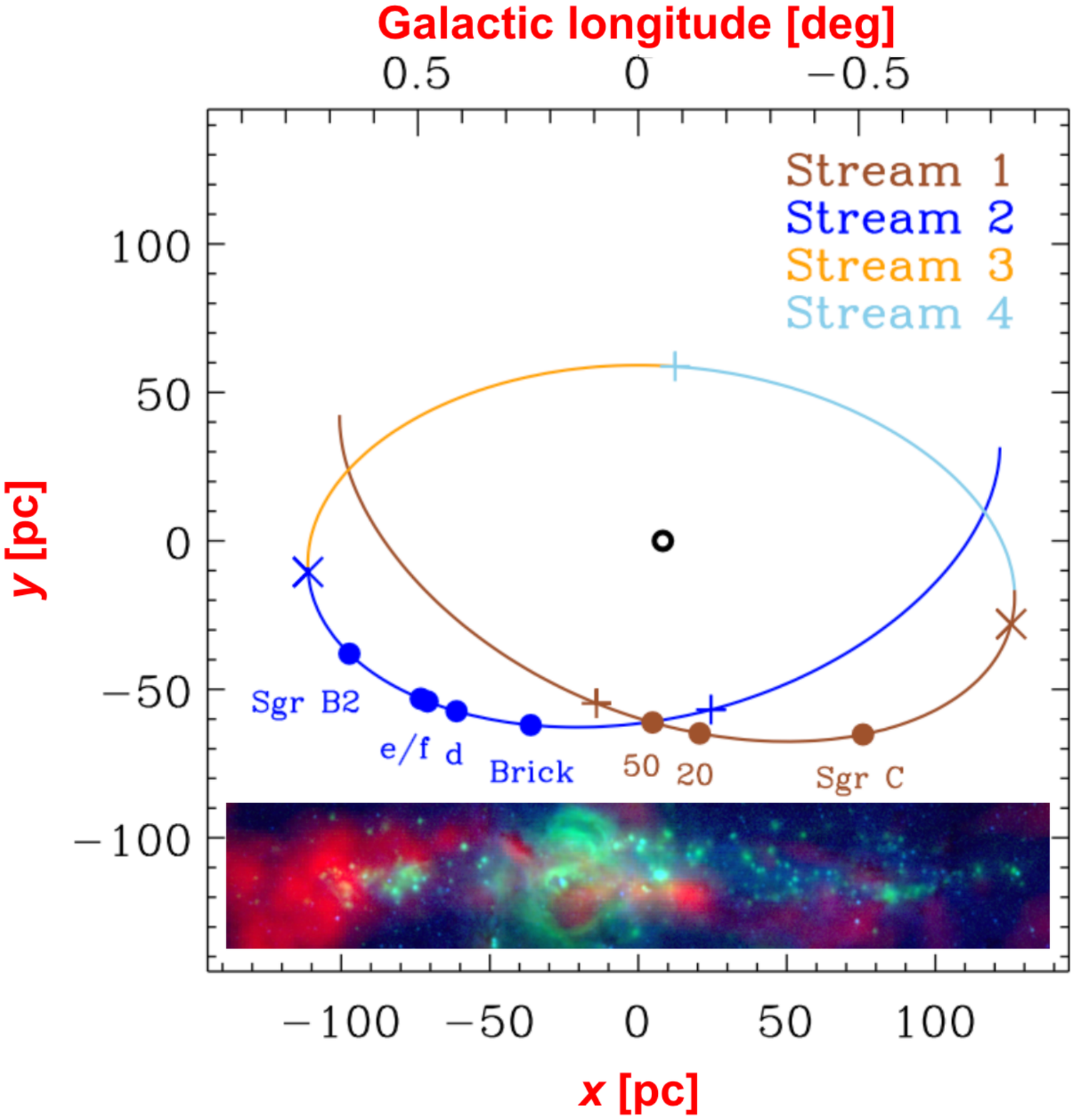}\hspace{0.5cm}
\includegraphics[width=6.44cm]{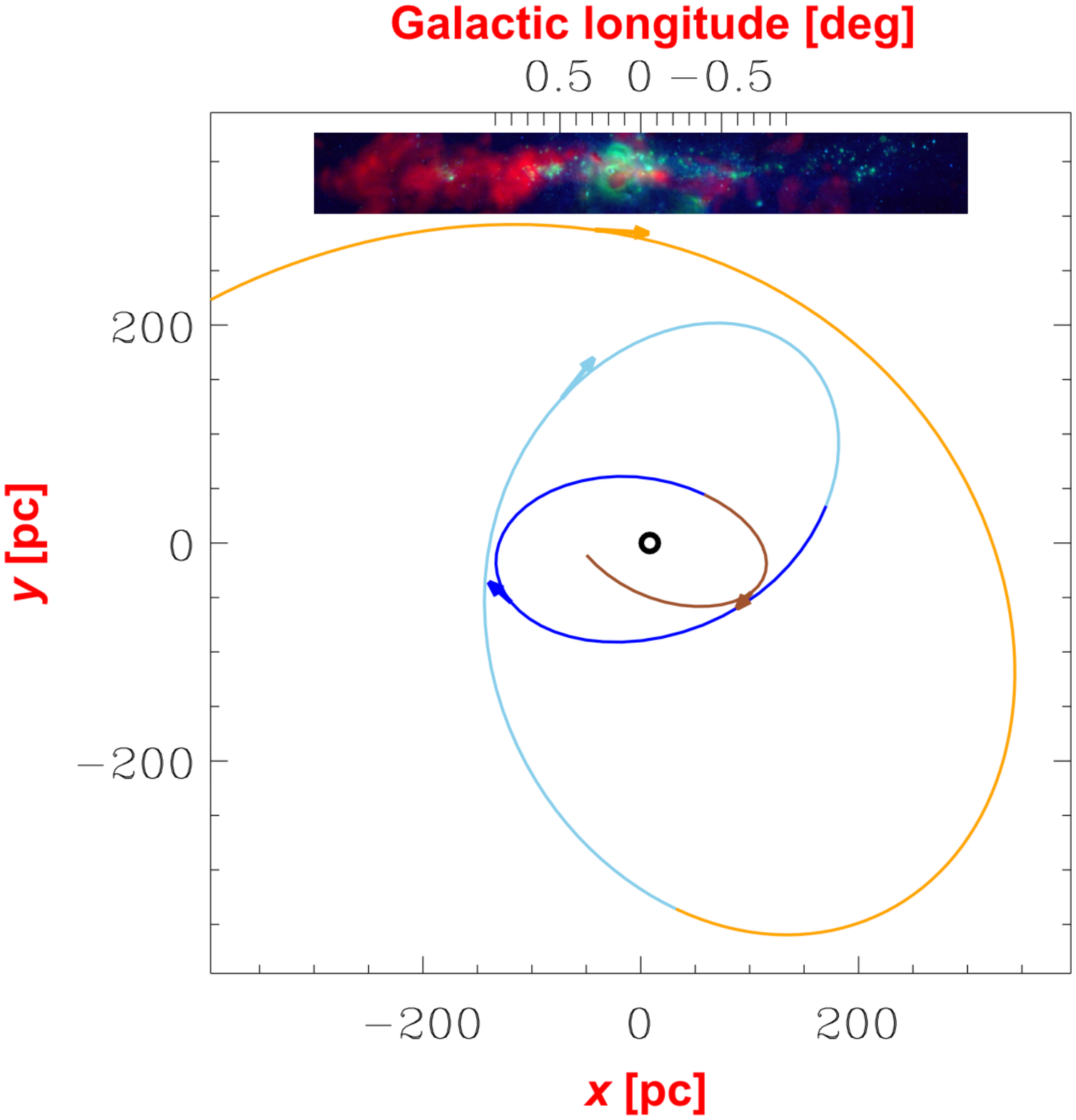}
\vspace{-1.5cm}
\caption{
\label{fig:orbit}
Orbital models revealing the three-dimensional geometry of the major gas streams in the CMZ, with the observer located in the negative-$y$ direction. {\it Left}: Top-down view of the best-fitting orbital solution to the 100-pc gas stream \citep[adapted from Figure~6 in][]{kruijssen15}. Several of the main clouds are indicated, as well as the pericentre ($+$) and apocentre ($\times$) passages, while the inset shows an NH$_3$(1,1)--21.3$\mu$m--8.3$\mu$m (red, green, blue) composite image of the same region as seen from Earth \citep[adapted from Figure~1 in][]{kruijssen14b}. The figure clearly shows that the orbit is not closed and that the most prominent CMZ clouds reside on the near side of the Galactic Centre. {\it Right}: The same orbital model with the addition of a constant radial velocity component of $-20$~km~s$^{-1}$ (Kruijssen et al.~in prep.), illustrating how the gas may have been deposited on the 100-pc stream from larger radii through the angular momentum transport process discussed in \S\ref{sec:macro}. Arrows indicate the direction of motion.
}
\end{figure}

Figure~\ref{fig:orbit} (left) shows a top-down view of the best-fitting orbital model from \citet{kruijssen15}, which provides the three-dimensional geometry of the gas stream. The figure shows that the gas resides on an eccentric orbit that is open rather than closed. The orbit covers a radial range from $R=60$--$120$~pc, with azimuthal and radial periods of 3.7~Myr and 2.0~Myr, respectively. While the figure only shows the part of the orbit traced in $\sim6$~Myr around the current positions of the gas clouds, the orbit can be integrated beyond that range. Over time, the stream will continue in the clockwise direction to trace a rosetta-shaped structure, extending the orbit shown here from the tip of Stream~1 in the top left of the diagram.

The orbital solution is chosen to conserve angular momentum, because the model of \S\ref{sec:macro} predicts that the angular moment transport by shear has stalled in the 100-pc stream. However, the deposition of gas onto this stream is characterised by a non-zero radial drift velocity, i.e.~the angular momentum of the gas evolves as it approaches the 100-pc stream. A first illustration of this process is shown in Figure~\ref{fig:orbit} (right). This model takes the near-side pericentre upstream of the Brick (also known as G0.253+0.016) from the left-hand panel and integrates the orbit from that point in both directions, this time including a constant radial velocity component of $-20$~km~s$^{-1}$ as an approximate parameterisation of the angular momentum transport and the resulting radial inflow (see \S\ref{sec:model} and \citealt{krumholz15}). The figure shows that the gas gradually spirals in towards the 100-pc stream, undergoing multiple pericentre passages before it reaches the stalling radius.

\subsection{Comparison to previous work}
Previous representations of the orbital structure and three-dimensional geometry of the gas in the CMZ differ from the \citet{kruijssen15} model in a number of key aspects. For instance, \citet{sofue95} and \citet{sawada04} proposed that the clouds are situated on two point-symmetric spiral arms, with Sgr~C and the 20 and 50~km~s$^{-1}$ clouds on the far side of the Galactic Centre. \citet{molinari11} parameterised the 100-pc stream using a closed ellipse with a constant circular velocity and placed the 20 and 50~km~s$^{-1}$ clouds much closer to an off-centre Sgr~A$^*$, at $R<20$~pc rather than $R=60$--$70$~pc in the \citet{kruijssen15} model.

\citet{henshaw16} carries out a detailed kinematic comparison between these three representations. It is concluded that the spiral arm geometry may be indistinguishable from the \citet{kruijssen15} model in terms of the $\{l,b,v_{\rm los}\}$ kinematics, but is inconsistent with the fact that the 20 and 50~km~s$^{-1}$ clouds are seen in absorption at $70\mu$m, which places them on the near side of the Galactic Centre. \citet{henshaw16} also finds that the closed ellipse provides a poor match to the observed kinematics of the 100-pc stream. This confirms the analysis of \citet{kruijssen15}, where it was shown that the shallow gradient of the line-of-sight velocity with Galactic longitude rules out geometries in which the 20 and 50~km~s$^{-1}$ clouds reside at radii $R\leq40$~pc. Further tests of these geometries will be carried out measuring the proper motions of masers within the clouds on the 100-pc stream using the Very Long Baseline Array (VLBA, see \S\ref{sec:outlook}).

\section{An evolutionary sequence of star-forming clouds}
\subsection{Cloud condensation and tidally-triggered collapse}
We now take a closer look at how star formation proceeds once the gas becomes prone to gravitational instability. As shown in Figure~\ref{fig:orbit} (right), the gas undergoes multiple pericentre passages while spiralling in towards the 100-pc stream. The tidal field during these passages is strongly compressive, especially in the vertical direction, and the passages therefore drive tidal perturbations of increasing strength as the gas spirals in. The growth time of the gravitational instabilities in the model of \citet{krumholz15} initially exceeds the radial oscillation period of the orbit, implying that the compressive tidal perturbations may play an important role in driving the condensation of self-gravitating molecular clouds, possibly giving them their final nudge into collapse.

The idea of tidally-triggered collapse of molecular clouds in the 100-pc stream was first proposed by \citet{longmore13b}, who noted that the clouds on the `dust ridge' (from the Brick to Sgr~B2 in Figure~\ref{fig:orbit}) exhibit an increase of the star formation efficiency downstream from the Brick. They hypothesised that this systematic behaviour results from tidally-triggered collapse during the preceding pericentre passage. If true, this would imply that the CMZ provides an extremely powerful probe of the physics of star formation and feedback, because in the orbital model of \citet{kruijssen15}, the time since pericentre passage (i.e.~$t=0$) is known. In other words, it would be possible to study these physics as a function of absolute time.

\begin{figure}
\center
\vspace{-0.4cm}
\includegraphics[width=10cm, angle=-90]{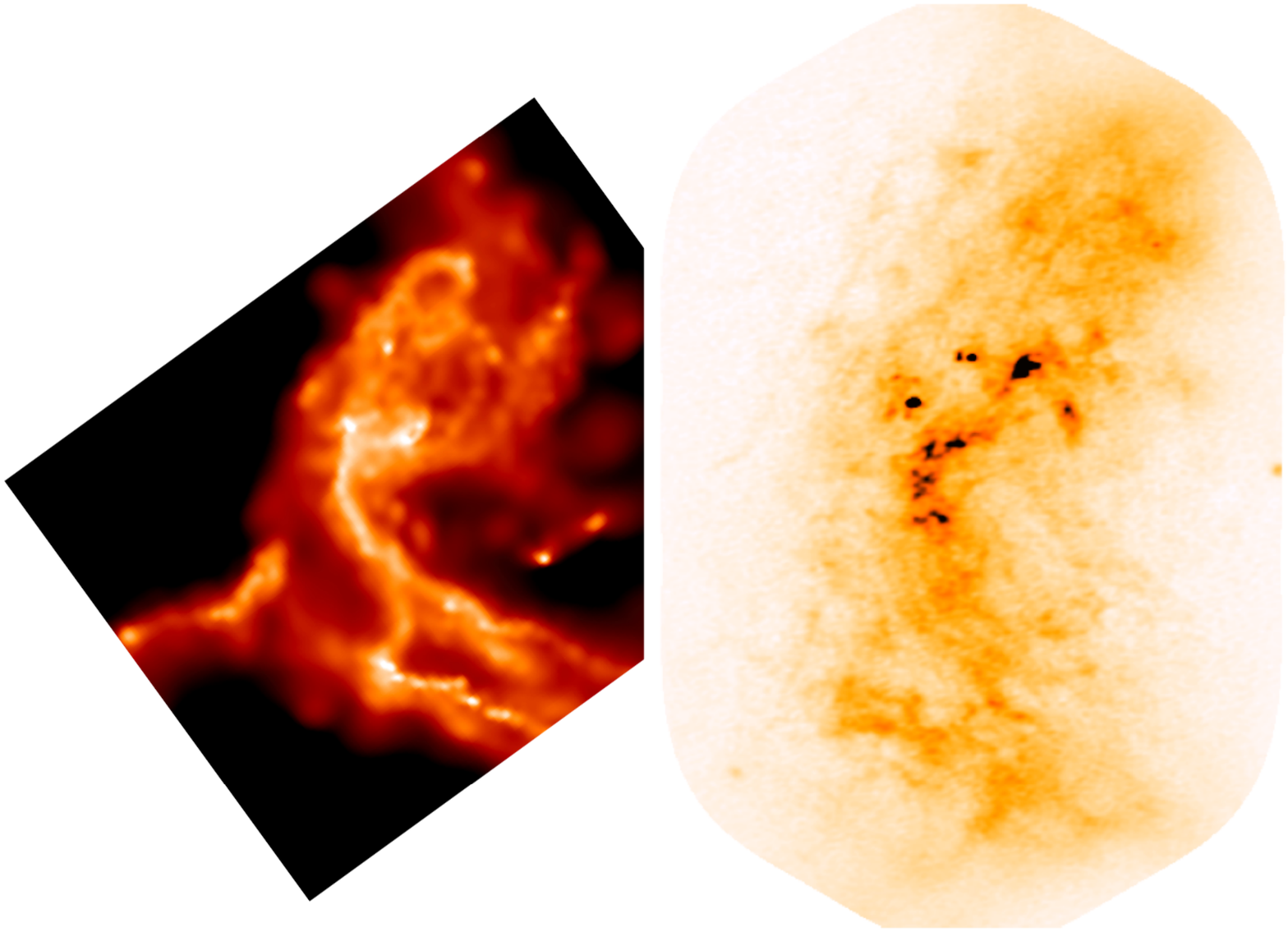}
\vspace{-0.3cm}
\caption{
\label{fig:brick}
Comparison between a hydrodynamical simulation of a Brick-like cloud on the \citet{kruijssen15} orbital model, showing a snapshot close to the position of the real Brick (left, Kruijssen, Dale, Longmore et al.~in prep.), and the ALMA dust image of the Brick (right, \citealt{rathborne15}). The simulation only focuses on modelling the cloud dynamics and it provides a good match to the observed curvature, the velocity gradient (not shown here), and even the structure of the cloud. As shown in Kruijssen, Dale, Longmore et al.~in prep., these observables are set by the cloud's response to the tidal deformation during the recent pericentre passage.
}
\end{figure}

To further investigate this scenario and assess the feasibility of an absolute evolutionary timeline, we are currently carrying out hydrodynamical simulations of molecular clouds on the \citet{kruijssen15} orbital model (Kruijssen, Dale, Longmore et al.~in prep.). One of these simulations is shown in Figure~\ref{fig:brick} (left) and follows a large ($r=20$~pc), turbulent ($\sigma_{1{\rm D}}=12$~km~s$^{-1}$), and massive ($M=2\times10^6$~M$_\odot$) gas reservoir on its evolution towards, through, and past pericentre. The simulation only includes hydrodynamics, self-gravity, initial turbulence, and sink particle formation. It is intended to focus on the influence of the orbital dynamics on the evolution of gas condensations forming out of the gas reservoir, and therefore excludes stellar feedback, magnetic fields, or turbulence driving. In this setup, the turbulent energy always dissipates, and the gas inevitably reaches a high star formation efficiency. However, we find that the sink particle formation rate is moderately accelerated relative to a control run of the same cloud on a circular orbit. Figure~\ref{fig:brick} shows a comparison between a single snapshot from this simulation close to the position of the Brick and the Atacama Large Millimeter/submillimeter Array (ALMA) dust image of the Brick \citep{rathborne15}. Despite the limited range of physics included in the simulation, it reproduces several of the key properties of the Brick, such as its curvature, its velocity gradient, and internal structure. While the latter is sensitive to the initial turbulent velocity field and therefore is likely a coincidence, the former two observables are set by the cloud's response to the tidal deformation during the recent pericentre passage (Kruijssen, Dale, Longmore et al.~in prep.). These simulations therefore support the idea that the Brick formed through tidally-triggered collapse.

\subsection{Evidence and implications}
Since the paper by \citet{longmore13b}, the observational evidence in favour of the tidally-triggered collapse scenario has also been mounting. \citet{barnes16} extend the gradual increase of star formation activity to an analogous increase of feedback activity even further downstream -- hot gas bubbles driven by the young massive stellar populations are systematically larger the further downstream from the pre-Brick pericentre passage they are. \citet{rathborne14} argue that the kinematics of the Brick are also consistent with tidally-triggered collapse, as it has a dynamical time-scale similar to the time since pericentre and exhibits bulk radial motions indicative of the expected tidal deformation. In addition, the density probability distribution function (PDF) of the Brick follows the traditional lognormal shape caused by turbulent motion, but the power law tail at high densities ( that indicates gravitational collapse and fragmentation towards star formation is scarcely populated \citep{rathborne14b,federrath16}. Because the cloud is gravitationally bound \citep{walker15}, this means that it must be young -- of the order of a dynamical time, which coincides with the time since the most recent pericentre passage. \citet{kauffmann13} and \citet{mills15} confirm the nearly starless nature of the Brick, further underlining its youth. Temperature measurements using line emission from H$_2$CO \citep{ginsburg16} and NH$_3$ (N.~Krieger et al., these proceedings) show that the dust ridge clouds exhibit a pronounced gradient of increasing temperature with time since pericentre passage. \citet{ginsburg16} show that this behaviour is quantitatively consistent with increased heating by turbulent energy dissipation, which is expected if the most recent pericentre passage nudged the clouds into gravitational collapse. 

In summary, there are clear indications of systematic cloud evolution with time since pericentre. However, if the gas condensations in the 100-pc stream upstream from pericentre have strongly differing properties (e.g.~densities, dynamical times, virial ratios), then there is no guarantee that they will all respond in the same way to the tidal perturbation during the upcoming pericentre passage. At the very least, this introduces noise in any absolute evolutionary timeline after pericentre, but in the worst case it may erase the systematic trends in evolutionary state altogether.

In \citet{henshaw16b}, we address this problem by studying the properties of the gas upstream from the pericentre passage near the Brick. We identify line-of-sight velocity corrugations as a function of Galactic longitude in this upstream gas, with an amplitude of $\sim4$~km~s$^{-1}$ and a wavelength of $\sim20$~pc. At the velocity extremes, the gas stream hosts compact ($R\sim2$~pc) condensations with masses of $\sim10^4$~M$_\odot$, that are quasi-regularly spaced with a separation of $\sim10$~pc. The wavelength and separation closely match the predicted Toomre length and Jeans length, showing that gravitational instabilities drive the initial condensation of molecular clouds from the gas stream (as predicted by models, see \S\ref{sec:macro}). This suggests that the seeds for the dust ridge clouds had similar properties. The tidally-triggered collapse of these condensations during their pericentre passage is therefore likely to proceed at similar rate, although individual outliers may exist due to small but present differences in density between the condensations -- the standard deviation of the $\{$masses, radii, volume densities, free-fall times$\}$ in the sample of condensations from \citet{henshaw16b} is $\{0.19, 0.09, 0.16, 0.08\}$~dex.

In view of the above differences, the `absolute timeline' of the post-pericentre clouds and star-forming regions is predicted to exist in the sense of a systematic trend, but is not expected to be strictly monotonic. To provide insight in the physics of star formation, the timeline should be expressed in terms of the number of elapsed free-fall times. The standard deviation of the free-fall time thus implies that the timeline itself has an uncertainty of 0.08~dex or 20\%. Such deviations likely increase further when comparing clouds near the different pericentre passages in the orbital model of Figure~\ref{fig:orbit}. It is therefore inadvisable to stack the cloud samples around each pericentre to improve statistics. Even if the trends downstream from other pericentre passages are similar to those seen on the dust ridge (e.g.~N.~Butterfield et al., these proceedings), systematic offsets will cause these trends to weaken significantly or vanish when stacked.

If the evidence in favour of an evolutionary sequence of star-forming clouds holds even after further and more detailed scrutiny, then the 100-pc stream represents a long-awaited, real-Universe analogy to a numerical simulation, in which correlated snapshots can be compared and followed, allowing the characterisation of cloud evolution, star formation, and feedback as a function of absolute time. Under this assumption, it is interesting to note that the quiescent Brick cloud and the vigorously star-forming Sgr~B2 complex are separated by $1$--$1.5$ free-fall times (i.e.~$0.4$--$0.5$~Myr) of evolution, indicating that star formation proceeds rapidly once collapse is initiated, even in the notoriously inefficiently star-forming CMZ (also see \citealt{kruijssen15} and \citealt{barnes16}). This simple example demonstrates the possible implications of the proposed scenario.

\section{Massive star and cluster formation in the CMZ}
At the high gas densities and pressures that characterise the CMZ in general and the dust ridge in particular, about 50\% of all star formation is predicted to take place in bound stellar clusters, which is about 5--10 times higher than in the solar neighbourhood \citep{kruijssen12d}. This high bound cluster formation efficiency is reached due to the high densities and short free-fall times in the CMZ, which can cause density peaks within clouds to locally exhaust their gas before feedback expels the residual gas or shuts off the gas inflow \citep{kruijssen12,longmore14,dale15,ginsburg16b}. The maximum mass scale of these clusters is thought to be set by the Toomre mass \citep{kruijssen14c,adamo15b}, i.e.~the largest scale that can collapse against shear. In the 100-pc stream, the Toomre mass is $2\times10^5$~M$_\odot$ \citep[cf.][]{henshaw16b}, which for the above 50\% cluster formation efficiency and a fiducial star formation efficiency of 10\% implies that the most massive clusters expected to form in the CMZ have masses of $\sim10^4$~M$_\odot$. This almost exactly matches the masses of the Arches and Quintuplet clusters \citep{portegieszwart10}, whereas the Toomre mass itself matches the mass of the Brick \citep{longmore12}. It is thus not guaranteed that the dust ridge clouds form Arches-like clusters, but it is certainly possible.

The Arches and Quintuplet clusters have ages and three-dimensional velocity vectors that are consistent with having formed on the dust ridge after accounting for their subsequent orbital motion \citep{kruijssen15}. Next to lending further support to the tidally-triggered collapse model, this also suggests that the dust ridge clouds may indeed be able to form Arches-like clusters. For the particular case of the Brick, this has been suggested in a large number of recent papers \citep[e.g.][]{longmore12,rathborne14,rathborne15,walker16b}. Even if the formation of Arches-like clusters is a common mode of star formation in the CMZ, it is important to note that such clusters are expected to be extremely short-lived, with disruption time-scales as low as $\sim10$~Myr \citep{kruijssen14b} due to a combination of tidal evaporation \citep{portegieszwart01} and impulsive shocks from passing gas clouds, which are prevalent in galactic centres \citep{kruijssen11}.

Even if the clusters do not survive, they represent an important bottleneck in the context of the global transport of gas towards Sgr~A$^*$. The gas mass fraction that gets locked up in stars is unlikely to add to further supermassive black hole growth. Conversely, feedback from Arches-like clusters may push gas onto highly eccentric orbits plunging towards Sgr~A$^*$, in which case star and cluster formation could stimulate black hole activity \citep[as has been suggested for extragalactic nuclei, see e.g.][]{davies07}.

\section{Outlook} \label{sec:outlook}
The CMZ provides a unique laboratory for studying star formation in an environment that is the best local analogue for the conditions seen in high-redshift galaxies. At the same time, it represents a hallmark example of a system in which large-scale gas flows, galactic dynamics, star formation, feedback, and the feeding of the central supermassive black hole are intertwined. While an understanding of the multi-scale physics driving the gas-star formation cycle in this system is slowly emerging, the number of open questions and exciting new avenues for follow-up work keeps increasing. In this context, it is promising that several large observational surveys of the CMZ are currently under way, with a clear focus on testing theoretical models. For instance, the CMZoom Survey with the Submillimeter Array (C.~Battersby et al., these proceedings) is providing a complete census of the massive protostellar cores in the CMZ, testing the current ideas on where and under which conditions (massive) star formation can commence in the CMZ. The SWAG survey (J.~Ott et al., these proceedings) is carrying out multi-line observations of the CMZ in H$_2$O and NH$_3$, enabling detailed temperature measurements and maser detections. Orbital models are about to be undergo a critical test thanks to an ongoing VLBA survey (Immer et al., in prep.), aiming to measure proper motions of masers in several clouds on the 100-pc stream. Finally, ALMA is now capable of observing extragalactic CMZs at resolutions similar to pre-ALMA observations of the Galactic CMZ.

Above all, the various Galactic Centre communities are starting to connect on key interdisciplinary questions in which CMZ research can play an important role, such as relating the gas-star formation cycle to the accretion activity of Sgr~A$^*$, as well as combining the constraints on the three-dimensional geometry of the gas from X-ray light echoes and orbital modelling. This will enable more accurate models of the absorption along the line of sight, which may potentially contribute to the search for dark matter annihilation signals. The Galactic Centre ecosystem may be one of the most complex physical systems in the local Universe, but the field is entering a very promising phase.

\section*{Acknowledgements}
The conference organisers are gratefully acknowledged for the kind invitation and for financial support, as well as for organising a very lively and stimulating conference. In addition, I am indebted to my collaborators for their insights and contributions to the projects discussed here. I particularly would like to thank Jim Dale, Jonathan Henshaw, Mark Krumholz, Steve Longmore, and Jill Rathborne. I gratefully acknowledge support in the form of an Emmy Noether Research Group from the Deutsche Forschungsgemeinschaft (DFG), grant number KR4801/1-1.

\bibliographystyle{iau}
\bibliography{mybib}

%\begin{discussion}
%
%\discuss{To be}{inserted by}
%
%\discuss{the}{Editors}
%\end{discussion}

\end{document}